\documentclass[12pt]{iopart}
\usepackage{iopams}

\begin{document}

\title{Alternative structures and bi-Hamiltonian systems on a Hilbert space}

\author{G Marmo\dag, G Scolarici\ddag, A Simoni\dag \footnote[3]{To
whom correspondence should be addressed (simoni@na.infn.it)} and F
Ventriglia\dag}

\address{\dag\ Dipartimento di Scienze Fisiche e Sezione INFN di Napoli, Universit\`{a} Federico II,
 Complesso Universitario Monte S. Angelo, via Cinthia, 80126 Naples, Italy}

\address{\ddag\ Dipartimento di Fisica e Sezione INFN di Lecce, Universit\`{a} di Lecce,
 via Provinciale per Arnesano, 73100 Lecce, Italy}

\begin{abstract}
We discuss transformations generated by dynamical quantum systems
which are bi-unitary, i.e. unitary with respect to a pair of
Hermitian structures on an infinite-dimensional complex Hilbert
space. We introduce the notion of Hermitian structures in generic
relative position. We provide few necessary and sufficient
conditions for two Hermitian structures to be in generic relative
position to better illustrate the relevance of this notion. The
group of bi-unitary transformations is considered in both the
generic and non-generic case. Finally, we generalize the analysis to
real Hilbert spaces and extend to infinite dimensions results
already available in the framework of finite-dimensional linear
bi-Hamiltonian systems.

\end{abstract}



\maketitle



\section{Introduction}
The general structures ruling the evolution of classical and quantum
systems are not essentially different. For instance both systems are
Hamiltonian vector fields and both are derivations on the Lie
algebra of observables with respect to the Poisson bracket and the
commutator bracket respectively. Besides, in some appropriate limit,
quantum mechanics should reproduce classical mechanics.\cite{dirac}
So the question arises of which alternative quantum descriptions for
a given quantum system would reproduce the alternative classical
descriptions of Hamiltonian systems.These systems are usually known
as bi-Hamiltonian systems. Completely integrable systems are often
associated with alternative compatible Poisson structures. We recall
that by compatibility is usually understood that any combination,
with real coefficients, of the two Poisson brackets satisfies the
Jacobi identity. In this respect, we should remark that while on a
vector space the imaginary part of the hermitian structures, i.e.
constant symplectic structures, are always mutually compatible, this
is not true for the full hermitian structures. In this case the
compatibility of the complex structures gives non trivial conditions
even in the vector space situation. As a matter of fact the complex
structure, related to the indetermination relation, plays no role in
the classical limit of quantum mechanics.\cite{bedlevo}

In the study of bi-Hamiltonian systems one usually starts with a
given dynamics and looks for alternative Hamiltonian descriptions
(see a partial list of references  for classical \cite{ma} and
 for quantum \cite{blo} systems).

In this paper we deal with a kind of converse problem \cite{msv},
i.e. we start with two Hermitian structures on a complex Hilbert
space and look for all dynamical quantum evolutions which turn out
to be bi-unitary with respect to them. This study generalizes our
previous results on finite-dimensional bi-Hamiltonian systems in
reference \cite{mor} to the infinite-dimensional case.

This paper is organized as follows. In section 2, we consider two
Hermitian structures on a finite-dimensional Hilbert space and show
the equivalence of the following three properties for the Hermitian
positive operator $G$ which connects them: the non-degeneracy, the
cyclicity and the genericity. A short description of a bi-unitary
group is also given. In section 3, we introduce the
infinite-dimensional case recalling the direct integral
decomposition of a Hilbert space with respect to a commutative ring
of operators, which is a suitable mathematical tool to deal with
such a situation \cite{nai}. In section 4, we extend to the
infinite-dimensional Hilbert spaces the analysis drawn in section 2.
In particular, we prove that the component spaces in the
decomposition are one-dimensional if and only if the Hermitian
structures are in relative generic position. Also, we show that this
happens if and only if the operator $G$ connecting the two Hermitian
structures is cyclic. This allows to conclude that all the quantum
systems, which are bi-unitary with respect to two Hermitian
structures in generic relative position, commute among themselves.
Moreover, the bi-unitary group is explicitly exhibited both in the
generic and non generic case. In section 5, the analysis starts from
different complexifications of a real Hilbert space to discuss the
previous results in the light of the notion of compatible
triples.\cite{mor, dasilva} In section 6 we discuss a simple example
of some physical interest and finally, in the last section, we draw
a few conclusions.

\section{Bi-unitary group on a finite-dimensional space}

In quantum mechanics the Hilbert space $\mathcal{H}$ is given as a \emph{%
complex} vector space, because the complex structure enters directly
the Schroedinger equation of motion.

Denoting with $h_{1}(.,.)$ and $h_{2}(.,.)$ two Hermitian structures
given on $\mathcal{H}$\ (both linear, for instance, in the second
factor), we search for the group of transformations which leave both
$h_{1}$ and $h_{2}$ invariant, that is the bi-unitary transformation
group.

By using the Riesz's theorem a bounded, positive operator $G$ may be
defined, which is self-adjoint both with respect to $h_{1}$ and
$h_{2}$, as:

\begin{equation}
 h_{2}(x,y)=h_{1}(Gx,y),\ \ \ \ \forall x,y\in \mathcal{H}.
\end{equation}

Moreover, any bi-unitary transformation $U$ must commute with $G$.
Indeed:

\begin{equation*}
\fl h_{1}(x,U^{\dagger
}GUy)=h_{1}(Ux,GUy)=h_{2}(Ux,Uy)=h_{2}(x,y)=h_{1}(Gx,y)=h_{1}(x,Gy)
\end{equation*}
and from this

\begin{equation}
U^{\dagger }GU=G \Leftrightarrow [G,U]=0.
\end{equation}
Therefore the group of bi-unitary transformations is contained in the commutant $%
G^{\prime }$ of the operator $G$.

To visualize these transformations, let us consider the bi-unitary
group of transformations when $\mathcal{H}$ is finite-dimensional.
In this case $G$ is diagonalizable and the two Hermitian structures
result proportional in each eigenspace of $G$ \emph{via} the
eigenvalue. Then the group of bi-unitary transformations is given by

\begin{equation}
U(n_{1})\times U(n_{2})\times ...\times U(n_{m}), \ \ \ %
n_{1}+n_{2}+...+n_{m}=n=\dim \mathcal{H},
\end{equation}
where $n_{k}$ denotes the degeneracy of the $k$-th eigenvalue of
$G$.

The picture should be clear now. Each Hermitian structure on
$\mathcal{H}$ defines a different realization of the unitary group
as a group of transformations. The intersection of these two groups
identifies the group of bi-unitary transformations.

In finite-dimensional complex Hilbert spaces the following
definition can be introduced \cite{mor}:

\noindent \textbf{Definition 1 }\textit{Two Hermitian forms are said
to be in generic relative position when the eigenvalues of
}$G$\textit{\ are non-degenerate.}

Then, if \ $h_{1}$ and $h_{2}$ are in generic position, the group of
bi-unitary transformations becomes

\begin{eqnarray*}
&&\underbrace{U(1)\times U(1)\times ...\times U(1)}. \nonumber\\
&&\ \ \ \ \ \ \ \ \ \ n\ \  factors \nonumber
\end{eqnarray*}

In other words, this means that $G$ generates a complete set of
commuting observables.

Now, recalling that an operator is cyclic when a vector $x_{0}$
exists such
that the set $\{x_{0},$ $Gx_{0},...,$ $G^{n-1}x_{0}\}$ spans the whole $n-$%
dimensional Hilbert space, we show that:

\noindent \textbf{Proposition 1}
 \textit{Two Hermitian forms are in generic relative position if and only if
their connecting operator }$G$\textit{\ is cyclic}.

\textbf{Proof }The non singular operator $G$ has a discrete spectrum
and is diagonalizable so, when $h_{1}$ and $h_{2}$ are in generic
position, $G$ admits $n$ distinct eigenvalues $\lambda _{k}$. Let
now $\{e_{k}\}$ be the eigenvector basis of $G$ and $\{\mu ^{k}\}$
an $n$-tuple of nonzero complex numbers. The vector
\begin{equation}
x_{0}=\sum\nolimits_k\mu ^{k}e_{k}
\end{equation}
is a cyclic vector for $G$. In fact one obtains
\begin{equation}
G^{m}x_{0}=\sum\nolimits_k\mu ^{k}\lambda _{k}^{m}e_{k}\ ,\ \ \
m=0,1,...,n-1.
\end{equation}
The vectors $\{G^{m}x_{0}\}$ are linearly independent because the
determinant of their components is given by
\begin{equation}
(\prod\limits_{k}\mu ^{k})V(\lambda _{1},...,\lambda _{n}),
\end{equation}
where $V$ denotes the Vandermonde determinant which is different
from zero when all the eigenvalues $\lambda _{k}$ are distinct. The
converse is also true.$\ \ \Box$

This shows that definition $(1)$ may be equivalently formulated as:

\noindent\textbf{Definition 2} \textit{Two Hermitian forms are said
to be in generic relative position when their connecting operator
}$G$\textit{\ is cyclic.}

The genericity condition can also be restated in a purely algebraic
form as follows:

\noindent \textbf{Definition 3} \textit{Two Hermitian forms are said
to be in generic relative position when }$G^{\prime \prime
}=G^{\prime }$\textit{, i.e. when the bi-commutant of }$G$
\textit{coincides with the commutant of} $G$.

Equivalence of definitions $(3)$ and $(1)$ is apparent.

The last two equivalent properties of $G$ are readily suitable for
an extension of the genericity condition to the infinite-dimensional
case while, at a first glance, the definition based on
non-degeneracy of the spectrum of $G$ looks hardly generalizable.

\section{Decomposing an infinite-dimensional Hilbert space}

Now we deal with the infinite-dimensional case, when the connecting
operator $G$ may have a point part and a continuum part in its
spectrum.

As regards to the point part, the bi-unitary group is
$U(n_{1})\times ...\times U(n_{k})\times ...,$ where now $n_{k}$ may
also be $\infty .$ When $G$ admits a continuum spectrum, the
characterization of the bi-unitary group is more involved and
suitable mathematical tools are needed from the spectral theory of
operators and the theory of rings of operators on Hilbert spaces.

We recall that each commutative (weakly closed) ring of operators
$C$ in a Hilbert space, containing the identity, corresponds to a
direct integral of Hilbert spaces.

The following theorems \cite{nai} are useful:

\noindent \textbf{Theorem 1 }\textit{To each direct integral of
Hilbert spaces with respect to a measure }$\sigma $\textit{\ on a
real interval }$\Delta :$

\begin{equation*}
\mathcal{H}=\int_{\Delta }H_{\lambda }\textrm{d}\sigma (\lambda ),
\end{equation*}
\textit{there corresponds a commutative weakly closed ring
}$C=L_{\sigma }^{\infty }(\Delta ),$\textit{\ where to each
}$\varphi \in L_{\sigma }^{\infty }(\Delta )$\textit{\ there
corresponds the operator }$L_{\varphi }:(L_{\varphi }\xi )=\varphi
(\lambda )\xi _{\lambda }$ \textit{with} $\xi
\in \mathcal{H},$ $\xi _{\lambda }\in H_{\lambda }$\textit{\ and }$%
||L_{\varphi }||=||\varphi ||_{\infty }.$
\bigskip

\emph{Vice versa}:

\bigskip

\noindent \textbf{Theorem 2 }\textit{To each commutative weakly closed ring }$C$%
\textit{\ of operators in a Hilbert space }$\mathcal{H}$\textit{\
there corresponds a decomposition of }$\mathcal{H}$\textit{\ into a
direct
integral, for which }$C$\textit{\ is the set of operators of the form }$%
L_{\varphi },$ $\varphi \in L^{\infty }$\textit{.}

To apply the previous theorems to the ring $R(G)$ generated by the
connecting operator $G$, we preliminarily remark that:

\noindent \textbf{Proposition 2} \textit{The weakly closed commutative ring }$R(G)$%
\textit{\ generated by the connecting operator }$G$ \textit{contains
the identity.}

\textbf{Proof} Let $E_{0}$ be the principal identity of $G$ in the ring of all bounded operators $\mathcal{B}(\mathcal{H}%
) :$ by definition $E_{0}$ is the projection operator on the
orthogonal complement of the set $\ker G.$

We recall \cite{nai} that the minimal weakly closed ring $R(G)$ containing $%
G $ contains only those elements $A\in G^{\prime \prime }$ which
satisfy, like $G,$ the following condition:
\begin{equation}
E_{0}A=AE_{0}=A.
\end{equation}

Now the positiveness of the operator $G$ ensures that $\ker G=0.$ This implies that $E_{0}=%
\mathbf{1}\in R(G).\ \ \  \square $

Then, by theorem (2), the ring $R(G)$ induces a decomposition of the Hilbert space $%
\mathcal{H}$ into the direct integral

\begin{equation}
\mathcal{H}=\int_{\Delta }H_{\lambda }\textrm{d}\sigma (\lambda ),
\label{Hilbert decomposition}
\end{equation}
where $\Delta =[a,b]$ contains the entire spectrum of the positive
self-adjoint operator $G.$ The measure $\sigma (\lambda )$ in
equation (\ref{Hilbert decomposition}) is obtained by the spectral
family $\{P_{G}(\lambda )\}$ of $G$ and cyclic vectors in the usual
way.\cite{nai}

We remark that it results $R(G)\equiv G^{\prime \prime }$. Therefore $%
G^{\prime \prime }$ is commutative.

Now every operator $A$ from the commutant $G^{\prime }$ is
representable in the form of a direct integral of operators

\begin{equation}
A\ \cdot=\int_{\Delta }A(\lambda )\ \cdot\ \textrm{d}\sigma (\lambda
),
\end{equation}
where $A(\lambda )$ is a bounded operator in $H_{\lambda }$, for
almost every $\lambda \in \Delta $.

Thus the bi-unitary transformations, as they belong to $G^{\prime}
,$ are in general a direct integral of unitary operators $U(\lambda
)$ acting on $H_{\lambda }$.

 In particular, every operator
$B$ of the bi-commutant $G^{\prime \prime }=R(G)$ is a
multiplication by a number $b(\lambda )$ on $H_{\lambda },$ for
almost every $\lambda :$
\begin{equation}
B(\lambda )= b(\lambda )\ 1_{\lambda} .
\end{equation}

\section{Bi-unitary group on an infinite-dimensional Hilbert space}

More insight can be gained from a more specific analysis of the
direct integral decomposition of $\mathcal{H}\mathbb{\ },$ which can
be written as

\begin{equation}
\mathcal{H}=\int_{\Delta }H_{\lambda }\textrm{d}\sigma (\lambda
)=\bigoplus\limits_{k}\int_{\Delta _{k}}H_{\lambda }\textrm{d}\sigma
(\lambda )=\bigoplus\limits_{k}\mathcal{H}_{k},
\label{hilbertdecompfine}
\end{equation}
where now the spectrum $\Delta $ of $G$ is the union of a countable
number of measurable sets $\Delta _{k}$, such that for $\lambda \in
\Delta _{k}$ the spaces $H_{\lambda }$ have the same dimension
$n_{k}$ ($n_{k}$ may be $\infty $).

The measure $\sigma (\lambda )$ is obtained by the measures $\sigma
_{k}(\lambda )$'s \textit{via } the spectral family $\{P_{G}(\lambda
)\}$ of $G$\ and cyclic vectors $u_{k}$ , with $\sigma _{k}(\lambda
)=(P_{G}(\lambda )u_{k},u_{k})$.

The dimension $n_{k}$ of the spaces $H_{\lambda }$ is the analog of
the degeneracy of the eigenvalues $\lambda $ of the point part of
the spectrum of $G$ .

According to the decomposition of equation
(\ref{hilbertdecompfine}), any operator $A$ in the commutant
$G^{\prime }$ is representable as:

\begin{equation}
A\ \cdot=\bigoplus\limits_{k}\int_{\Delta _{k}}A(\lambda ) \ \cdot\
\textrm{d}\sigma (\lambda ).  \label{operatordecomp}
\end{equation}

In particular, the connecting operator $G$ is a multiplication by
$\lambda $ on each $H_{\lambda }$, so we get the following result at
once:

\noindent\textbf{Proposition 3 }\textit{Let two Hermitian structures} $h_{1}$ \textit{%
and} $h_{2}$ \textit{be given on the Hilbert space
}$\mathcal{H}$\textit{. Then there exists a decomposition of
}$\mathcal{H}$ \textit{into a direct
integral of Hilbert spaces }$H_{\lambda }$\textit{\ such that in each space }%
$H_{\lambda }$\textit{\ the structures\ }$h_{1}|_{H_{\lambda }}$
\textit{and} $h_{2}|_{H_{\lambda }}$\textit{\ are proportional:
}$h_{2}|_{H_{\lambda }}=\lambda \ h_{1}|_{H_{\lambda }}$\textit{.}

Moreover, as $G$ acts like a multiplicative operator on each
component space $H_{\lambda },$ the expressions of $h_{1}$ and
$h_{2}$ on $\mathcal{H}$ are:

\begin{equation*}
h_{1}(x,y)=\sum_{k}\int\nolimits_{\Delta
_{k}}<x_{\lambda},y_{\lambda}>_{\lambda}\textrm{d}\sigma (\lambda )\
\ ,
\end{equation*}
\begin{equation}
h_{2}(x,y)=\sum_{k}\int\nolimits_{\Delta _{k}}\lambda
<x_{\lambda},y_{\lambda}>_{\lambda }\textrm{d}\sigma (\lambda )
\label{inner}
\end{equation}
where $<x_{\lambda},y_{\lambda}>_{\lambda }$ is the inner product on
the component $H_{\lambda }$.

As a consequence of proposition (3) and equation
(\ref{operatordecomp}), the elements $U$ of the bi-unitary group
acting on $\mathcal{H}$ have the form:

\begin{equation}
U\ \cdot=\bigoplus\limits_{k}\int_{\Delta _{k}}U_{n_{k}}(\lambda )\
\cdot \ \textrm{d}\sigma (\lambda ),  \label{unitdecomp}
\end{equation}
where $U_{n_{k}}(\lambda )$ is an element of the unitary group
$U(n_k) $ for each $\lambda \in \Delta _{k}.$

As regards to the notion of two Hermitian forms in generic position,
the following statement \cite{lecce} holds:

\noindent\textbf{Proposition 4}\textit{\ Two Hermitian structures }$h_{1}$ \textit{and%
} $h_{2}$\ \textit{are in generic relative position if and only if
the component
spaces }$H_{\lambda }$\textit{\ of the decomposition of }$\mathcal{H}$%
\textit{\ into a direct integral\ with respect to }$R(G)$
\textit{are one-dimensional. }

\textbf{Proof} Let us suppose that two Hermitian forms are given in
generic relative position. Then, by definition (3), $R(G)=G^{\prime \prime }=G^{\prime }$, so $%
G^{\prime }$ is commutative and any component operator $A(\lambda )$
in
equation (\ref{operatordecomp}) acts on an one-dimensional component space $%
H_{\lambda }$, for almost every $\lambda \in \Delta $.

In order to prove the converse, observe that if $R(G)=G^{\prime
\prime }\neq G^{\prime }$, then $G^{\prime }$ is not commutative. So
a subset $\Delta _{0}$ of $\Delta $ exists such that $\dim
H_{\lambda }>1$ for $\lambda \in \Delta _{0}.\ \ \ \square $

This shows the equivalence of definitions (1) and (3) also in the
infinite-dimensional case.

Propositions (3) and (4) extend to infinite-dimensional complex
Hilbert spaces some results of our previous work \cite{mor}, so that
we can say that all quantum dynamical bi-Hamiltonian systems are
pairwise commuting if (and only if) the two Hermitian structures are
in generic relative position.

In the generic case, the unitary component operators
$U_{n_k}(\lambda )$ in
equation (\ref{unitdecomp}) reduce to a multiplication by a phase factor $%
\textrm{exp}(\textrm{i} \vartheta (\lambda ))$ on $H_{\lambda }$ for
almost every $\lambda $, so that the elements of the bi-unitary
group read

\begin{equation}
U\ \cdot=\int_{\Delta }\rm{e}^{\rm{i}\vartheta (\lambda )}\ \cdot \
\textrm{d}\sigma (\lambda ).
\end{equation}

Therefore in the generic case the group of bi-unitary
transformations is parameterized by the $\sigma -$measurable real
functions $\vartheta $ on $\Delta .$ This shows that the bi-unitary
group may be written as
\begin{equation}
U_{\vartheta }=\textrm{exp}(\textrm{i}\vartheta (G))\; .
\end{equation}

Finally, like in the finite-dimensional case, an equivalence may be
stated between the genericity condition and the cyclicity of the
operator $G$. In fact, we have:

\noindent\textbf{Proposition 5\ }\textit{Let }$G$\textit{\ be a
bounded positive self-adjoint operator in }$\mathcal{H}.$\textit{
Then }$G$\ \textit{is cyclic if and only if }$G^{\prime \prime
}=G^{\prime }.$

\textbf{Proof} Let us suppose $G^{\prime \prime }=G^{\prime }$. Then
$R(G)=G^{\prime \prime }=G^{\prime }$ and $G^{\prime }$ is
commutative. Hence the decomposition of the Hilbert space yields
one-dimensional component
spaces $H_{\lambda }$ where $G$ acts as a multiplication by $\lambda $ in $%
L_{2}(\Delta ,\sigma ).$ Then the vector $x_{0}=1/\lambda $ is a
cyclic vector in $L_{2}(\Delta ,\sigma )$, so $G$ is cyclic.

Conversely, let $G$ be cyclic. Then each space $H_{\lambda }$ is
one-dimensional and any operator from $G^{\prime }$ acts as a
multiplication by a number in $H_{\lambda }$. Hence $G^{\prime
}=R(G)=G^{\prime \prime }.\ \ \  \square $

Summarizing, we have shown the equivalence of definitions (1), (2)
and (3) in the infinite-dimensional case.

\section{Compatible structures on a real infinite-dimensional Hilbert space }

In the previous section we have analyzed the setting of a complex
Hilbert
space $\mathcal{H}$ with two Hermitian structures\ $h_{1}(.,.)$ and $%
h_{2}(.,.)$ and now, to make contact with real linear Hamiltonian
mechanics \cite{mor} on infinite dimensional spaces, we analyze the
consequences of this on real Hilbert spaces. Besides, the real
context\ displays richer contents and is a more general setting for
the analysis of our geometric structures.

We start therefore with a real vector space
$\mathcal{H}^{\mathcal{R}}$ (isomorphic to the realification of
$\mathcal{H}$).
From the two Hermitian structures on the previous complex Hilbert space, $%
h_{1}(.,.)$ and $h_{2}(.,.),$\ we get on $\mathcal{H}^{\mathcal{R}}$
two metric tensors $g_{a}$ and two symplectic forms $\omega _{a}$
\textit{via }: 
\begin{equation*}
g_{a}(x,y)=\Re \ h_{a}(x,y);\ \ \omega _{a}(x,y)=\Im \ h_{a}(x,y)\ ,
\ a=1,2.
\end{equation*}
On $\mathcal{H}^{\mathcal{R}}$
the multiplication by
the imaginary unit appears as the action of a linear operator $J$ , $%
J^{2}=-1,$ which is skew-adjoint with respect to both $g$'s.

The structures are related by the equation $\omega
_{a}(x,y)=g_{a}(Jx,y)$ which defines the \emph{admissible} triples
$(g_{a},\omega _{a},J)$.

Then the three linear operators $G^{\mathcal{R}}=g_{1}^{-1}\circ
g_{2},T=\omega _{1}^{-1}\circ \omega _{2}=-J\circ
G^{\mathcal{R}}\circ J$ and $J$ are a set of mutually commuting
linear operators, $G^{\mathcal{R}}$ and $T$ being self-adjoint with
respect to both metric tensors. We remark, by the way, that $T$ is
the recursion operator for symplectic structures.

For instance, to check that $[G^{\mathcal{R}},J]=0,$ consider the
equation $h_{2}(x,y)= h_{1}(G x,y)$ which defines the connecting
operator $G.$ Then:
\begin{eqnarray}
\fl h_{1}(G x,y) =g_{1}(G x,y)+\textrm{i}g_{1}(J G x,y)=h_{2}(x,y) \nonumber\\
=g_{2}(x,y)+\textrm{i}g_{2}(J
x,y)=g_{1}(G^{\mathcal{R}}x,y)+\textrm{i}g_{1}(G^{\mathcal{R}}J
x,y). \nonumber
\end{eqnarray}
This shows, by equating real and imaginary parts, that $G^{\mathcal{R}}=G$ and $%
[G,J]=0 .$ It is trivial now that $[T,G]=[T,J]=0$ as well. By
definition this means that these two triples are
\emph{compatible}.\cite{mor}

Quantum theory in the usual complex context leads quite naturally to
consider identical complex structures in the two triples. On the
contrary, in the real context it is possible to consider the case of
two distinct complex
structures $J_{1},J_{2}$. In other words, on a real Hilbert space $\mathcal{H}%
^{\mathcal{R}}$ let two admissible triples $(g_{1},J_{1},\omega _{1})$ and $%
(g_{2},J_{2},\omega _{2})$  be given which are compatible, that is
the commuting operators $\left\{ G,T,J_{1},J_{2}\right\} $ have the
correct bi-Hermiticity properties.\cite{dasilva}

Now it is possible to complexify $\mathcal{H}^{\mathcal{R}}$ and to
get a complex Hilbert space $\mathcal{H}_{1}$ with a Hermitian
scalar
product $<.,.>_{1}\ $ \textit{ via }$\ (g_{1},J_{1},\omega _{1}).$ Since by hypothesis the operators $%
\left\{ G,T,J_{2}\right\} $ commute with $J_{1}$, they become
complex-linear operators on $\mathcal{H}_{1}$. In particular $G$
becomes a complex-linear bounded positive self-adjoint operator,
therefore $G$ acts as a multiplication by $\lambda $ on the
component spaces in the associated direct integral decomposition

\begin{equation}
\mathcal{H}=\int_{\Delta }H_{\lambda }\textrm{d}\sigma (\lambda ).
\end{equation}
Now $J_{2}$ commutes with $G$, i.e. $J_{2}\in G^{^{\prime }}$ , so
$J_{2}$
is block-diagonal on $\mathcal{H}$. In each $H_{\lambda },$ we have $%
J_{2}^{2}(\lambda )=-1_{\lambda }$ and $J_{2}^{\dagger }(\lambda
)=-J_{2}(\lambda )$ . Then $H_{\lambda }$ splits in two parts
corresponding to the eigenvalues $\pm $i of $J_{2}(\lambda ):
H_{\lambda }=H_{\lambda }^{+}\oplus H_{\lambda }^{-},$ where on
$H_{\lambda }^{+}:J_{2}=J_{1}=$i, while on $H_{\lambda
}^{-}:J_{2}=-J_{1}=-$i. The direct integral decomposition becomes:

\begin{equation}
\fl \mathcal{H}=\int_{\Delta }H_{\lambda }^{+}\oplus H_{\lambda
}^{-}\;\textrm{d}\sigma (\lambda )=\mathcal{H}^{+}\oplus
\mathcal{H}^{-}=\int_{\Delta ^{+}}H_{\lambda }^{+}\;
\textrm{d}\sigma (\lambda )\oplus \int_{\Delta ^{-}}H_{\lambda
}^{-}\;\textrm{d}\sigma (\lambda ), \label{pmdecomp}
\end{equation}
where $\Delta ^{+}$ and $\Delta ^{-}$, subsets of $\Delta $ not
necessarily disjoint, are support of  $H_{\lambda }^{+}$ and
$H_{\lambda }^{-}$ respectively. This completely extends the
finite-dimensional result in \cite{mor}.

At this point we can draw a complete picture: starting from two
admissible triples $(g_{a},J_{a},\omega _{a}),\ a=1,2,$ on
$\mathcal{H}^{\mathcal{R}}$ we may construct the corresponding
Hermitian structures $h_{a}=g_{a}+\textrm{i}\omega _{a}$. We stress
that $h_{a}$ is a Hermitian structure on $\mathcal{H}_{a},$ which is
the
complexification of $\mathcal{H}^{\mathcal{R}}\ $ \emph{via }$\ %
J_{a}$ , so that in general $h_{1}$ and $h_{2}$ are not Hermitian
structures on the $\emph{same}$ complex vector space.

When the triples are compatible
the decomposition of the space in equation (\ref{pmdecomp}) holds, so that $%
\mathcal{H}^{\mathcal{R}}$ can be decomposed into the direct sum of
the spaces $ \mathcal{H}_{\mathcal{R}}^{+}$ and
$\mathcal{H}_{\mathcal{R}}^{-}$
on which $J_{2}=\pm J_{1},$ respectively. The comparison of $h_{1}$ and $%
h_{2} $ requires a fixed complexification of
$\mathcal{H}^{\mathcal{R}}$, for instance
$\mathcal{H}_{1}=\mathcal{H}_{1}^{+}\oplus \mathcal{H}_{1}^{-}.$
Then, using equations (\ref{inner}) and (\ref{pmdecomp}), we can
write
\begin{equation}
h_{1}(x,y)=\int\nolimits_{\Delta
^{+}}<x_{\lambda},y_{\lambda}>_{\lambda}\textrm{d}\sigma (\lambda
)+\int\nolimits_{\Delta ^{-}}<x_{\lambda},y_{\lambda}>_{\lambda
}\textrm{d}\sigma (\lambda )\ \ ,
\end{equation}
while
\begin{equation}
h_{2}(x,y)=\int\nolimits_{\Delta ^{+}}\lambda
<x_{\lambda},y_{\lambda}>_{\lambda }\textrm{d}\sigma (\lambda
)+\int\nolimits_{\Delta ^{-}}\lambda
<y_{\lambda},x_{\lambda}>_{\lambda }\textrm{d}\sigma (\lambda )\ .
\label{scalarprod}
\end{equation}
It is apparent that $h_{2}$ is not a Hermitian structure as it is
neither linear nor anti-linear on the whole space $\mathcal{H}_{1}.$

\section{Example: Particle in a box, a double case}

Consider the operator $G=1+X^{2}$ , with $X$ position operator, on $%
L_{2}([-\alpha ,\alpha ],dx)$. It is Hermitian with spectrum $\Delta
=[1,1+\alpha ^{2}]$. From the spectral family of $X:$%
\begin{equation}
P(\lambda )f=\chi _{[-\alpha ,\lambda ]}f \ ,
\end{equation}
where $\chi _{[-\alpha ,\lambda ]}$ is the characteristic function
of
the interval $[-\alpha ,\lambda ]$, we get the spectral family $%
P_{G}(\lambda )$ of $G$:
\begin{equation}
P_{G}(\lambda )=P(\sqrt{\lambda -1})-P(-\sqrt{\lambda -1})\ .
\label{spectralfam}
\end{equation}
In fact, by a simple computation it is immediate to check that
$P_{G}$ is a projection operator:
\begin{equation}
P_{G}^{2}=P_{G},\ \ P_{G}(0)=0,\ \ P_{G}(\alpha ^{2})=1.
\end{equation}
Furthermore, write $G$ as
\begin{equation}
\fl G\ \cdot=\int\limits_{[-\alpha ,\alpha ]}(1+\lambda
^{2})\cdot\textrm{d} P(\lambda )=\int\limits_{[-\alpha
,0]}(1+\lambda ^{2})\cdot\textrm{d} P(\lambda
)+\int\limits_{[0,\alpha ]}(1+\lambda ^{2})\cdot\textrm{d} P(\lambda
)\ ,
\end{equation}
and change variable putting $\lambda =-\sqrt{\mu -1}$ in the first
integral and $\lambda =\sqrt{\mu -1}$ \ in the second one.
Eventually, the spectral decomposition of $G$ reads

\begin{equation}
G\ \cdot=\int\limits_{[1,1+\alpha ^{2}]}\lambda \ \cdot \ \textrm{d}
P_{G}(\lambda )\ ,
\end{equation}
where $P_{G}(\lambda )$ is given by equation (\ref{spectralfam}).

Now $G$ does not have cyclic vectors on the whole $L_{2}([-\alpha
,\alpha ],\textrm{d}x)$, because if $f$ is any vector, $xf(-x)$ is
non-zero and orthogonal to all powers $G^{n}f$. In other words
$G^{\prime}$, which contains both $X$ and the parity operator, is
not commutative.

This argument fails on $L_{2}([0,\alpha ],\textrm{d}x)$, where $\chi
_{[0,\alpha ]}$ is cyclic. Analogously, $\chi _{[-\alpha ,0]}$ is
cyclic on $L_{2}([-\alpha ,0],\textrm{d}x),$ so we get the splitting
in two $G$-cyclic spaces
\begin{equation}
L_{2}[-\alpha ,\alpha ]=L_{2}[-\alpha ,0]\oplus L_{2}[0,\alpha ]\ .
\end{equation}

From $P_{G}$ and those cyclic vectors we obtain the measure
\begin{equation}
\sigma (\lambda )=(P_{G}(\lambda )\chi _{\lbrack 0,\alpha ]},\chi
_{\lbrack 0,\alpha ]})=\sqrt{\lambda -1}
\end{equation}
for the decomposition of the Hilbert space
\begin{equation}
\mathcal{H}=\int\limits_{[1,1+\alpha ^{2}]}H_{\lambda }\ \textrm{d}\sigma (\lambda )%
\ ,
\end{equation}
where $H_{\lambda }$ is one-dimensional for the particle in the
$[0,\alpha ]$ box while is bi-dimensional for the $[-\alpha ,\alpha
]$ box.

The general case of an asymmetric box $[-\alpha ,\beta  ]$ is a
direct superposition of the two previous cases, as we have shown in
section 4: in fact, assuming $ \beta
>\alpha$ for instance, the decomposition becomes the direct sum of
 bi-dimensional spaces for the $[-\alpha ,\alpha ]$ box plus
 one-dimensional spaces for the  $[\alpha ,\beta  ]$ box.

The bi-unitary transformations $U$ read

\begin{equation}
U\ \cdot=\int\limits_{[1,1+\alpha
^{2}]}\textrm{e}^{\textrm{i}\varphi (\lambda )}\ \cdot \
\textrm{d}\sqrt{\lambda -1}
\end{equation}
in the $[0,\alpha ]$ box, and

\begin{equation}
U\ \cdot=\int\limits_{[1,1+\alpha ^{2}]}U_{2}(\lambda )\ \cdot \
\textrm{d}\sqrt{\lambda -1}
\end{equation}
in the $[-\alpha ,\alpha ]$ box. Finally, in the $[-\alpha ,\beta ]$
box:

\begin{equation}
U\ \cdot=\int\limits_{[1,1+\alpha ^{2}]}U_{2}(\lambda )\ \cdot \
\textrm{d}\sqrt{\lambda -1}\ \oplus\ \int\limits_{[1+\alpha
^{2},1+\beta ^{2}]}\textrm{e}^{\textrm{i}\varphi (\lambda )}\ \cdot
\ \textrm{d}\sqrt{\lambda -1}\ .
\end{equation}

\section{Concluding remarks}

In this paper we have shown how to extend to the more realistic case
of infinite dimensions the results of our previous paper dealing
mainly with finite level quantum systems. Our approach shows, in the
framework of quantum systems, how to deal with ``pencils of
compatible Hermitian structures''  in the same spirit of ``pencils
of compatible Poisson structures'' \cite{ge, imm}. We hope to be
able to extend these results to the evolutionary equations for
classical and quantum field theories.

\end{document}